\documentclass{jps-cp}
\usepackage{txfonts} %Please comment out this line unless the txfonts package is availabe in your LaTeX system.
\usepackage{url}
\usepackage{graphicx}
\usepackage{xfrac}
\usepackage{caption}

\title{Impact of Uncertainties in Astrophysical Reaction Rates on Nucleosynthesis in the $\nu p$ Process}

\author{T. \textsc{Rauscher}$^{1,2,3}$, N. \textsc{Nishimura}$^{4}$, G. \textsc{Cescutti}$^{5}$, R. \textsc{Hirschi}$^{3,6,7}$,
A. St.J. \textsc{Murphy}$^{3,8}$,\\
and C. \textsc{Fr\"ohlich}$^{9}$}

\inst{$^{1}$Department of Physics, University of Basel, 4056 Basel, Switzerland\\
$^{2}$Centre for Astrophysics Research, University of Hertfordshire, Hatfield AL10 9AB, United Kingdom\\
$^{3}$UK Network for Bridging Disciplines of Galactic Chemical Evolution (BRIDGCE), \url{https://www.bridgce.ac.uk}\\
$^{4}$Center for Gravitational Physics, Yukawa Institute for Theoretical Physics, Kyoto University, Kyoto 606-8502, Japan\\
$^5$INAF, Osservatorio Astronomico di Trieste, I-34131 Trieste, Italy\\
$^6$Astrophysics Group, Faculty of Natural Sciences, Keele University, Keele ST5 5BG, UK\\
$^7$Kavli IPMU (WPI), University of Tokyo, Kashiwa 277-8583, Japan\\
$^8$School of Physics and Astronomy, University of Edinburgh, Edinburgh, EH9 3FD, UK\\
$^9$Department of Physics, North Carolina State University, Raleigh, NC 27695-8202, USA}

\recdate{September 6, 2019}

\abst{The $\nu p$ process appears in proton-rich, hot matter which is expanding in a neutrino wind and may be realised in 
explosive environments such as core-collapse supernovae or in outflows from accretion disks. The impact of uncertainties in nuclear 
reaction cross sections on the finally produced abundances has been studied by applying Monte Carlo variation of all astrophysical 
reaction rates in a large reaction network. As the detailed astrophysical conditions of the $\nu p$ process still are unknown, a 
parameter study was performed, with 23 trajectories covering a large range of entropies and $Y_\mathrm{e}$. The 
resulting abundance uncertainties are given for each trajectory. The $\nu p$ process has been speculated to contribute to the light 
$p$ nuclides but it was not possible so far to reproduce the solar isotope ratios. It is found that it is possible to reproduce the 
solar $^{92}$Mo/$^{94}$Mo abundance ratio within nuclear uncertainties, even within a single trajectory. The solar values of the 
abundances in the Kr-Sr region relative to the Mo region, however, cannot be achieved within a single trajectory. They may still be 
obtained from a weighted superposition of different trajectories, though, depending on the actual conditions in the production 
site. For a stronger constraint of the required conditions, it would be necessary to reduce the uncertainties in the 3$\alpha$ and 
$^{56}$Ni(n,p)$^{56}$Co rates at temperatures $T>3$ GK.
}

\kword{nucleosynthesis, supernovae, $\nu p$-process, $p$-nuclides, nuclear reactions, Monte Carlo}

\begin{document}
\maketitle

\section{Introduction}
\label{sec:intro}

The Monte Carlo approach to simultaneously vary a large number of reaction rates in a nuclear reaction network has been shown to be 
a superior method to assess nuclear physics uncertainties in nucleosynthesis studies. The MC framework \textsc{PizBuin} 
\cite{tommy} is applicable in postprocessing of astrophysical trajectories with a large reaction network, 
accounting for several thousand nuclides with several tens of thousand reactions. The trajectories specify the temporal 
evolution of density and temperature and can be taken from any astrophysical simulation.  
\textsc{PizBuin} so far was already applied to a number of processes: the 
$\gamma$ process in core-collapse supernovae (ccSN) \cite{tommy}, 
the production of $p$ nuclei in white dwarfs exploding as 
thermonuclear (type Ia) supernovae (SNIa) \cite{snIa}, the weak $s$ process in massive stars \cite{nobuya}, and the main $s$ 
process in 
AGB stars \cite{gabriele}.

Here, we report on an uncertainty study for the $\nu p$ process \cite{mcnup}. In proton-rich, very dense matter at 
temperatures exceeding 2$-$3 GK, proton capture reactions ensue and establish (p,$\gamma$)-($\gamma$,p) equilibria within isotonic 
chains on the proton-rich side of the nuclear chart. In such an $rp$ process (rapid proton capture process) matter flows between 
neighbouring isotonic chains of nuclides occur via comparatively (relative to the process timescale) slow electron 
captures or $\beta^+$ decays \cite{schatz}. An acceleration of the $rp$ process is possible in the presence of electron 
anti-neutrinos, converting a small fraction of free protons to free neutrons by the charged current reaction 
$\overline{\nu}_\mathrm{e} + \mathrm{p} \rightarrow \mathrm{n} + \mathrm{e}^+$. This allows to connect isotonic chains by (n,p) 
reactions which are much faster than the weak reactions \cite{carla,carla_apj,pruet}. This leads to a significant production of 
proton-rich nuclei beyond Ni even at the short timescale of an explosive process. While it was initially proposed that ejecta from 
deep layers in ccSN would experience such conditions, the picture is less clear nowadays because recent ccSN simulations lack layers 
with the required combination of entropy and electron abundance $Y_\mathrm{e}$. Nevertheless, the possibility for a $\nu p$ process 
remains, either in ccSN or other sites, such as in matter outflows from accretion disks in merging events of neutron stars or black 
holes. This is why we performed a parameter study, investigating a number of distinct trajectories in hot, adiabatically expanding 
plasma 
and covering electron abundances of $0.55 \leq Y_\mathrm{e} \leq 0.725$ and entropies of $11.4\leq S\leq 184$ 
$k_{B}$~baryon$^{-1}$, taken as initial values at the time of freeze-out from nuclear statistical equilibrium (NSE) at 7 GK. The 
$Y_\mathrm{e}$ did not change significantly until the cessation of the $\nu p$ process below about 3 GK. 

\section{Monte Carlo variations}
\label{sec:mc}

Astrophysical reaction rates have to consider the thermal excitation of nuclei in the stellar plasma. They include reactions on 
target nuclei in the ground state (g.s.) and in excited states, depending on the plasma temperature. The higher the temperature, 
the more important reactions on excited states become. In the MC variations,
temperature-dependent rate uncertainties were used, determined from the relative contributions of reactions on the g.s.\ and on 
excited states at a given temperature. This is necessary because at stability the g.s.\ contributions to the stellar rate are 
often constrained experimentally and thus bear a smaller uncertainty than purely theoretical rate predictions. The total 
uncertainty factor of a reaction rate is obtained from a weighted sum of uncertainties for g.s.\ and excited states (see 
\cite{tommy,stellarerrors,sensi,advances} for details).
Each reaction is assigned its own uncertainty and a random multiplier drawn from the uncertainty range is used in each of the 10000 
MC runs used to generate a distribution of final abundances from a network calculation with the multiplied reaction rates. Although 
the MC multiplier is drawn from a uniform distribution of values within the uncertainty range, the values of the final abundance of 
each nuclide rather are lognormally distributed. This is typical for distributions stemming from a multiplication of uncertainties 
\cite{jaynes,mc1} as it is present in the combined action of many reactions contributing to the final abundance.

For the $\nu p$-nucleosynthesis study presented here it is important to note that the nucleosynthesis path is located a few units 
away from stability and therefore there are no experimentally determined reaction rates available (except for the 3$\alpha$ rate 
and a few reactions acting on stable nuclides at late times, see Section~\ref{sec:results}). Furthermore, the temperatures in the 
$\nu p$ process are so high that reactions on thermally excited states dominate the reaction rate \cite{sensi,advances}. 
Thus, the uncertainties in the reaction rates were dominated by the assumed theory uncertainties as specified in \cite{tommy}. For 
example, the two most important reaction types, (n,p)$\leftrightarrow$(p,n) and (p,$\gamma$)$\leftrightarrow$($\gamma$,p), were 
varied from $\sfrac{1}{3}$ the standard rate to twice the standard rate and (p,$\alpha$)$\leftrightarrow$($\alpha$,p) rates were 
varied between $\sfrac{1}{10}$ and twice the standard rate. It is worth noting that forward and reverse rates were always 
multiplied by the same factor in the MC variation because forward and reverse \textit{stellar} rates are connected by detailed 
balance.

In some cases only one or a few reactions dominate the uncertainty in the final abundance of a specific nuclide. This can be 
identified automatically by a strong correlation between the variation of the reaction rate and the variation in the final 
abundance. The correlations can be extracted from the stored MC data after all MC runs have been completed. The method is superior 
to visual inspection of flows and manual variation of limited rate sets, especially for identifying the most important reactions in 
complex flow patterns. A number of correlation definitions are found in literature. We chose the Pearson product-moment 
correlation coefficient as the most suitable and most easy to handle method to
quantify correlations between rates and abundances \cite{pearson,kendall55}. This correlation coefficient assumes values $0\leq 
|r|\leq 1$, where
positive values of the Pearson coefficients 
indicate a direct correlation between rate change and abundance change, and negative values signify an inverse correlation, 
i.e., the abundance decreases when the rate is increased. Larger values $|r|$ indicate a stronger correlation and in the current 
context we defined a key rate (i.e., a rate dominating the final uncertainty 
in the production of a nuclide)
by showing $|r|\geq 0.65$ in connection with the abundance of a specific nuclide.

In this parameter study of the $\nu p$ process, we used MC variation only for reactions on isotopes of Fe and heavier elements. 
Reactions on lighter nuclides were kept at their standard values taken from compilations. The 3$\alpha$ reaction is known to be a 
key reaction because it 
determines the $^{56}$Ni seed available for further processing up to heavier masses and thus regulates the effectivity of the $\nu 
p$ process. A variation of this rate strongly affects all final abundances and would cover the impact of other rate 
variations. Figure \ref{fig:f1} shows the difference in production of intermediate and heavy nuclei depending on the chosen 
3$\alpha$ rate. Trajectories with the same label also provide the same conditions. In 
\cite{mcnup} we explored this effect in more detail but chose to use the rate recommended by \cite{fynbo} for the MC variations of 
all other rates.
The reaction $^{56}$Ni(n,p)$^{56}$Co was also not varied together with all other rates because it plays a similar role in the $\nu 
p$ process as the 
3$\alpha$ reaction. Its variation also would mask the importance of any other rate. A more detailed discussion of its impact 
can be found in \cite{mcnup}.

\section{Results}
\label{sec:results}

\begin{figure}[tb]
\includegraphics[viewport=0 0 340 120,clip,width=\columnwidth]{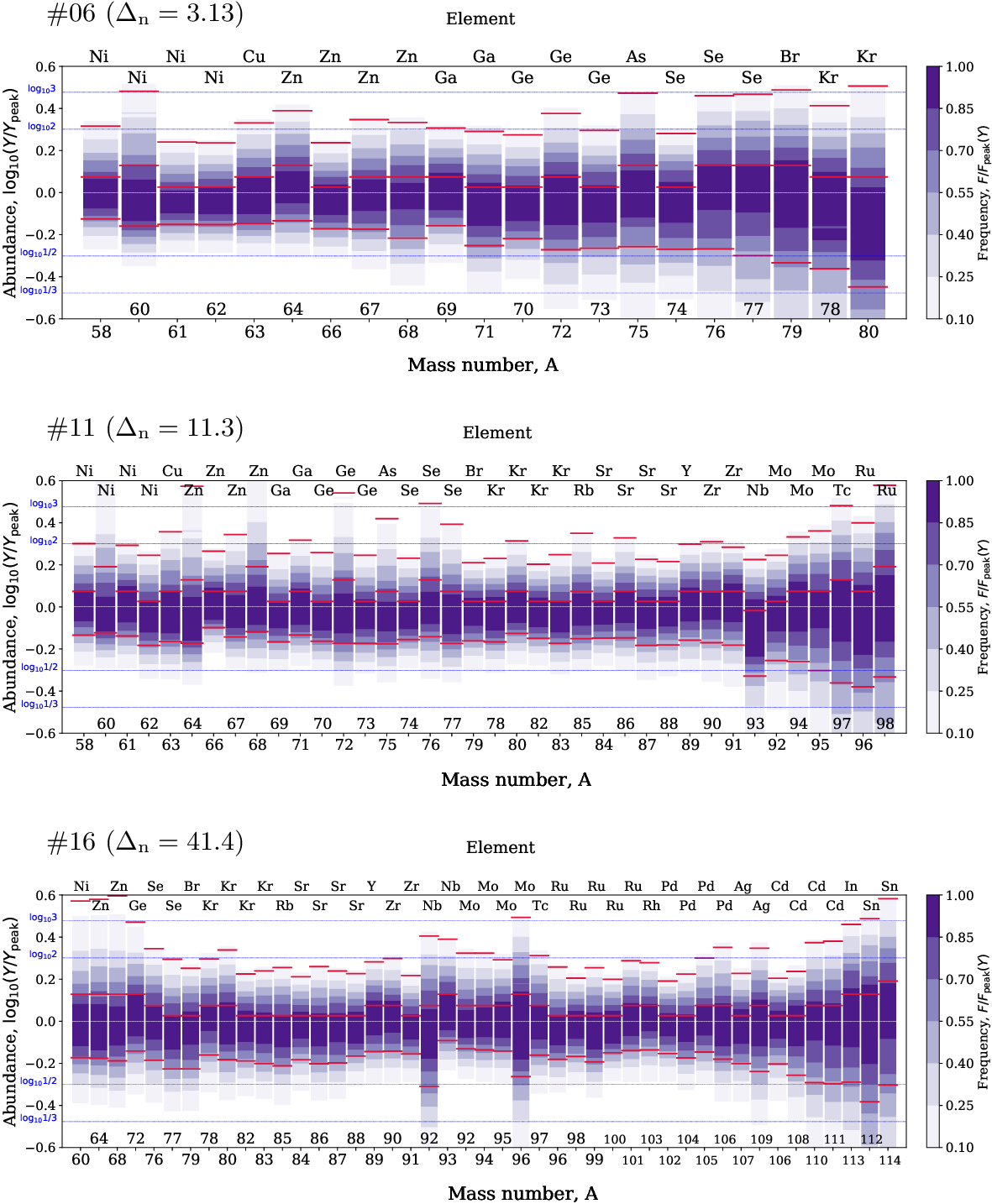}
\caption{Uncertainties in the final abundances caused by rate uncertainties for trajectory \#16. The two outer red lines in each 
distribution 
encompass a 90\% confidence interval. The third red line in between the outer red lines marks the 50\% summed probability. It does 
not necessarily coincide with the most probable abundance value at the darkest color shade, at which the distributions are 
centered. (Figure taken from \cite{mcnup}, with 
permission.)}
\label{fig:uncertall}
\end{figure}

\begin{figure}
\includegraphics[width=0.5\columnwidth]{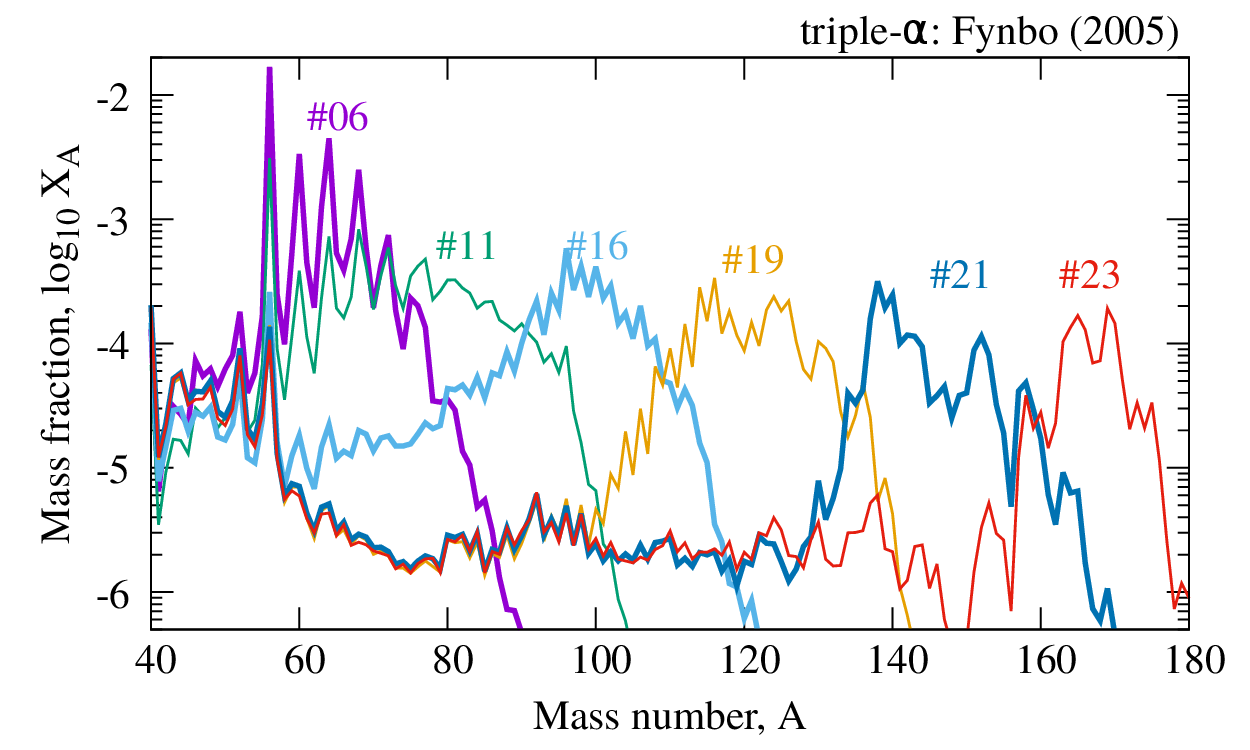}
\includegraphics[width=0.5\columnwidth]{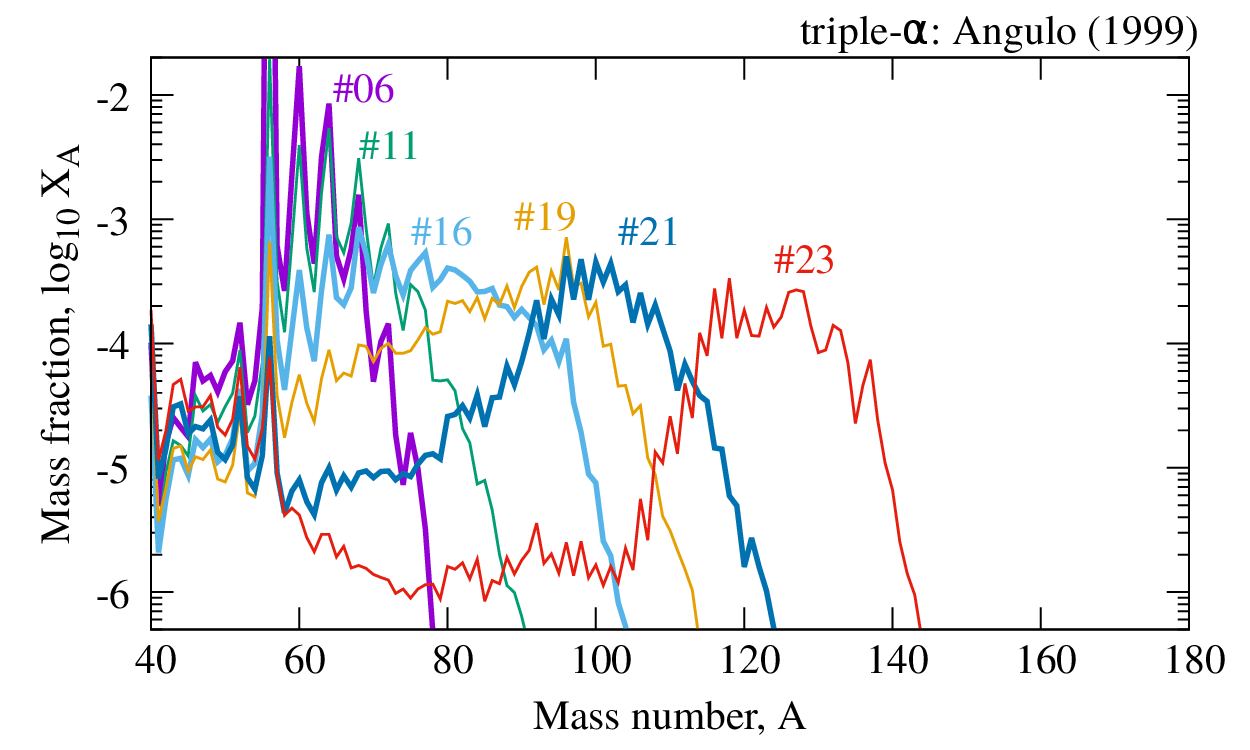}
\caption{Nucleosynthesis in selected trajectories; left panel: using the 
3$\alpha$ rate of Fynbo et al.\ \cite{fynbo}; right panel: using the 3$\alpha$ rate of NACRE \cite{nacre}. (Figures taken from 
\cite{mcnup}, with 
permission.)}
\label{fig:f1}
%\end{figure}
\medskip
%\begin{table}[tbh]
\captionof{table}{Isotope ratios and their uncertainties in selected trajectories using the 3$\alpha$ rate by Fynbo et 
al.\ \cite{fynbo} (see also Fig.\ 
\ref{fig:f1}, left panel).
\label{tab:ratio}}
\begin{center}
\begin{tabular}{c|cccc|cccc|cccc}
Traj.&\multicolumn{4}{c}{$Y(^{92}\mathrm{Mo})/Y(^{94}\mathrm{Mo})$}&\multicolumn{4}{c}{$Y(^{78}\mathrm{Kr})/Y(^{94}\mathrm{Mo})$}
&\multicolumn{4}{c}{$Y(^{84}\mathrm{Sr})/Y(^{94}\mathrm{Mo})$}\\
%Traj.&\multicolumn{4}{c}{(solar: 1.6)}&\multicolumn{4}{c}{(solar: 0.82)}&\multicolumn{4}{c}{(solar: 0.54)}\\
\#&$Y_{50}$&upper&lower&$Y_\mathrm{peak}$&$Y_{50}$&upper&lower&$Y_\mathrm{peak}$&$Y_{50}$&upper&lower&$Y_\mathrm{peak}$\\
\hline
11&1.2&2.1&0.8&0.9&2.8&3.6&0.5&2.2&2.4&3.0&0.6&1.9\\
16&0.8&2.8&0.7&0.6&0.1&2.8&0.6&0.1&0.3&2.5&0.6&0.2\\
17&       0.7&       3.3&       0.6&       0.5&       0.1&       2.3&       0.5&       0.1&       0.2&       2.7&       0.6&       
0.1\\
18&  0.8&  3.4&  0.6&  0.6&  0.2&  5.1&  0.6&  0.1&  0.3&  3.6&  0.5&  0.2\\
19&1.1&3.0&0.6&0.9&0.4&2.5&0.6&0.3&0.7&2.4&0.6&0.5\\
21&1.3&2.9&0.7&1.0&0.5&2.3&0.7&0.4&0.9&2.3&0.7&0.7\\
23&1.3&2.9&0.7&1.0&0.5&2.3&0.7&0.4&0.9&2.2&0.8&0.7\\
\hline
\end{tabular}
\end{center}
%\end{table}
\medskip
%\begin{table}[tbh]
\captionof{table}{Same as Table \ref{tab:ratio} but using the 3$\alpha$ rate by NACRE \cite{nacre} (see also Fig.\ 
\ref{fig:f1}, right panel). \label{tab:rationacre}}
\begin{center}
\begin{tabular}{c|cccc|cccc|cccc}
Traj.&\multicolumn{4}{c}{$Y(^{92}\mathrm{Mo})/Y(^{94}\mathrm{Mo})$}&\multicolumn{4}{c}{$Y(^{78}\mathrm{Kr})/Y(^{94}\mathrm{Mo})$}
&\multicolumn{4}{c}{$Y(^{84}\mathrm{Sr})/Y(^{94}\mathrm{Mo})$}\\
%Traj.&\multicolumn{4}{c}{(solar: 1.6)}&\multicolumn{4}{c}{(solar: 0.82)}&\multicolumn{4}{c}{(solar: 0.54)}\\
\#&$Y_{50}$&upper&lower&$Y_\mathrm{peak}$&$Y_{50}$&upper&lower&$Y_\mathrm{peak}$&$Y_{50}$&upper&lower&$Y_\mathrm{peak}$\\
\hline
11&  2.9&  1.8&  0.6&  2.7&813& 20&  0.5&284&117&  7.9&  0.5& 64\\
16&  1.2&  1.8&  0.6&  1.2&  2.7&  3.7&  0.6&  2.0&  2.5&  3.1&  0.6&  1.9\\
17&       1.1&       1.7&       0.6&       1.1&       1.3&       3.0&       0.6&       1.0&       1.5&       2.8&       0.7&       
1.1\\
18&  1.1&  1.8&  0.6&  1.0&  0.7&  2.1&  0.5&  0.6&  1.0&  2.0&  0.6&  1.0\\
19&  1.0&  1.9&  0.6&  1.0&  0.3&  2.0&  0.5&  0.3&  0.7&  1.9&  0.6&  0.7\\
21&  0.9&  3.1&  0.6&  0.7&  0.1&  3.0&  0.5&  0.1&  0.2&  2.8&  0.6&  0.2\\
23&  0.9&  3.4&  0.5&  0.6&  0.1&  9.3&  0.7&  0.0&  0.2&  4.9&  0.5&  0.2\\
\hline
\end{tabular}
\end{center}
%\end{table}
\end{figure}

Exemplary for the uncertainty distributions obtained in the MC variations, Fig.\ \ref{fig:uncertall} shows the results for 
trajectory \#16 using the 3$\alpha$ rate by \cite{fynbo}. This trajectory has its maximum production around mass number $A\simeq 
100$, as can be seen in Fig.\ 
\ref{fig:f1}. Its initial conditions were $Y_\mathrm{e}=0.65$ and $S=105$ $k_{B}$~baryon$^{-1}$ at 7 GK (see Sec.\ 
\ref{sec:intro}). Most of the final uncertainties are below a factor of two despite of the fact that the nucleosynthesis 
flow involves theoretical rates with larger uncertainties.

The $\nu p$ process is also interesting regarding the origin of certain proton-rich, unstable isotopes which cannot be synthesized 
by the $s$- and $r$-process, the so-called $p$ nuclei \cite{p-review}. Previous studies had problems to obtain the solar abundance 
ratio of $^{92}$Mo and $^{94}$Mo \cite{wanajo,xing}. Also for the relative abundance level of the Kr-Sr region and the Mo region, 
the solar value could not be obtained. Tables \ref{tab:ratio} and \ref{tab:rationacre} list the $^{92}$Mo/$^{94}$Mo, 
$^{78}$Kr/$^{94}$Mo, and 
$^{84}$Sr/$^{94}$Mo abundance ratios together with their uncertainties as obtained in our MC study \cite{tables}. The $Y_{50}$ 
value 
multiplied by factors "upper" and "lower" define an error interval enclosing 90\% of the probability distribution. ($Y_{50}$ and 
the upper and lower bounds are also marked by red lines in Fig.\ \ref{fig:uncertall}.) The most probable 
abundance value $Y_\mathrm{peak}$, according to the probability distribution (marked by the darkest color shade in Fig.\ 
\ref{fig:uncertall}), does not necessarily coincide with $Y_{50}$ because of the skewed distribution.

Table \ref{tab:ratio} clearly shows that when using the 3$\alpha$ rate by \cite{fynbo} all trajectories reproduce the solar 
$^{92}$Mo/$^{94}$Mo abundance ratio (1.6) within uncertainties. 
Trajectory \#16, however, most efficiently produces Mo.
The solar values of the $^{78}$Kr/$^{94}$Mo (0.82) and 
$^{84}$Sr/$^{94}$Mo (0.54) abundance ratios could be found concurrently at conditions close to those in trajectory \#18. This may 
be a spurious result, however, as all of the isotopes in question are barely produced in this trajectory and would be severely 
underproduced with 
respect to nuclei around mass numbers $A\simeq115-125$, see Fig.\ \ref{fig:f1}.

For the ratios listed in Table \ref{tab:rationacre} the 3$\alpha$ rate by \cite{nacre} was used. As is also demonstrated in Fig.\ 
\ref{fig:f1}, this 3$\alpha$ rate makes the $\nu p$ process less effective and suppresses the production of heavier nuclides 
compared to the production obtained with the rate by \cite{fynbo} under the same conditions otherwise. With this 3$\alpha$ rate 
the solar $^{92}$Mo/$^{94}$Mo abundance ratio can be reproduced within uncertainties in trajectories from \#12 up, where 
the flow through the progenitor chains of these nuclides is fully established. The solar value of the $^{84}$Sr/$^{94}$Mo 
abundance ratio is found in trajectories \#19 and up. The trajectories beyond \#19, however, strongly 
underproduce the Kr, Mo, Sr isotopes with respect to heavier elements, see Fig.\ \ref{fig:f1}. The spoiler for simultaneous 
co-production of the Kr, Mo, Sr isotopes within a single trajectory is the $^{78}$Kr/$^{94}$Mo ratio which only reproduces the 
solar 
value in trajectories \#17 (barely) and \#18 (best) when using the rate by \cite{nacre}. Only a very narrow window of conditions 
between \#18 and \#19 may allow concurrent reproduction of all ratios.

Further key rates beyond the 3$\alpha$ and $^{56}$Ni(n,p)$^{56}$Co rates were identified for each trajectory by using the 
correlation coefficient as described in Sec.\ \ref{sec:mc}. 
As expected, (n,p) reactions were dominant because they control the flow between equilibrated isotone chains. Several proton and 
neutron captures were also identified as key rates at the edge of the reaction network, impacting the (low-level) production of 
heavier nuclides. Extensive tables for key reactions affecting the uncertainties in the final abundances for a particular 
nuclide are given in \cite{mcnup}. Concerning the $^{92}$Mo/$^{94}$Mo abundance ratio, the reaction $^{92}$Mo(p,$\gamma$)$^{93}$Tc 
was identified as key reaction. It is inversely correlated with the ratio, implying that an increase in the rate leads to a decrease 
in the $^{92}$Mo/$^{94}$Mo abundance ratio. An experimental determination of this rate recently appeared after we commenced with 
our MC project \cite{mo92}. It only determines the ground state contribution to the stellar rate, however. At the elevated 
temperature of the $\nu p$ process, reactions on excited states of $^{92}$Mo significantly contribute to the rate. These are not 
directly constrained by existing experimental studies.

\section{Conclusion}

A comprehensive, large-scale MC study of nucleosynthesis in the $\nu p$ process has been performed. A range of conditions in a 
$Y_\mathrm{e}$ and entropy parameter-space was explored to cover the possibilities regarding implementations of nucleosynthesis in 
different sites. Our results allow to quantify the uncertainties stemming from nuclear physics input for any particular 
astrophysical simulation spanning this wide range of $Y_\mathrm{e}$ and entropy parameter-space.

For each of the chosen trajectories, the 
astrophysical reaction rates for several thousand target nuclides for Fe and above were simultaneously varied within individual 
temperature-dependent uncertainty ranges constructed from a combination of experimental and theoretical error bars. The 
combined effect of rate uncertainties was investigated, leading to total uncertainties in the final abundances of stable nuclei 
obtained after the $\nu p$ process had ceased. Key rates dominating the uncertainties in the final yields were determined. 
Different key rates were found for each trajectory as the production range of nuclides depends on the thermodynamic conditions.

Concerning the isotope ratios of light $p$ nuclides it was found that it is possible to reproduce the solar 
$^{92}$Mo/$^{94}$Mo abundance ratio within uncertainties, even though only rate uncertainties and not mass uncertainties 
have been considered. The reproduction of both the Mo isotopic ratio and their production level relative to the lighter $p$ isotopes 
of Kr and Sr has been found to be difficult within a single trajectory, regardless of whether the 3$\alpha$ rate by 
\cite{fynbo} or by \cite{nacre} is used. It may still be conceivable that a 3$\alpha$ rate between the two choices could bring all 
ratios into accordance with solar values for conditions close to those in our trajectories \#18 or \#19 \cite{mcnup}.
Stronger conclusions on whether this is
actually feasible are pending further reduction in the uncertainties of the 3$\alpha$ and $^{56}$Ni(n,p)$^{56}$Co rates at high 
temperature ($>3$ GK).
Moreover, a
parameter study like the present investigation 
is not devised to address a superposition of conditions, as may arise in a realistic description. In a realistic nucleosynthesis 
site a range of conditions covering several of our trajectories with different weights, may be realized. Such a superposition would 
also potentially allow different ratios of isotopic abundances.

In summary, we found that the uncertainties in the production of nuclei are dominated by the uncertainties arising from the choice 
of site, explosion model, and numerical treatment of the explosion hydrodynamics, as these crucially determine what range of 
nuclei can actually be produced. With the exception of the 3$\alpha$ and $^{56}$Ni(n,p)$^{56}$Co rates, which strongly 
affect the efficiency of nucleosynthesis in the $\nu p$ process, reaction rate uncertainties give rise to final abundance 
uncertainties in the range of factors of only $2-3$.

\section*{Acknowledgments}
This work has been partially supported by the European Research Council (EU-FP7-ERC-2012-St Grant 306901, EU-FP7 Adv Grant 
GA321263-FISH), the EU COST Action CA16117 (ChETEC), the UK STFC (ST/M000958/1), and MEXT Japan. G.C. acknowledges financial support 
from the EU Horizon2020 programme under the Marie
Sk\l odowska-Curie grant 664931. C.F. acknowledges support by the U.S.\ DOE under Award No.~DE-FG02-02ER41216. Computations 
were carried out partially on COSMOS (STFC DiRAC Facility) at DAMTP of the 
University of Cambridge, funded by BIS National E-infrastructure capital grant ST/J005673/1, STFC capital grant 
ST/H008586/1, and STFC DiRAC Operations grant ST/K00333X/1. DiRAC is part of the UK National E-Infrastructure. Further computations 
were carried out at CfCA, NAO Japan, and YITP, Kyoto University. The University of Edinburgh is 
a charitable body, registered in Scotland, with Registration No.~SC005336.

\end{document}